\input{aipcheck}

\documentclass[
    ,final            % use final for the camera ready runs
  ]
  {aipproc}

\layoutstyle{6x9}

\begin{document}

\title{Complex-mass scheme and resonances in EFT}


\author{T. Bauer}{
  address={Institut f\"ur Kernphysik, Johannes
Gutenberg-Universit\"at, D-55099 Mainz, Germany
}}

\author{D. Djukanovic}{
  address={Helmholtz Institut Mainz, D-55099 Mainz, Germany}
}

\author{J. Gegelia}{
  address={Institut f\"ur Kernphysik, Johannes
Gutenberg-Universit\"at, D-55099 Mainz, Germany},
altaddress={High
Energy Physics Institute of TSU, 0186 Tbilisi, Georgia}
}

\author{S. Scherer}{
  address={Institut f\"ur Kernphysik, Johannes
Gutenberg-Universit\"at, D-55099 Mainz, Germany}}

\author{L. Tiator}{
  address={Institut f\"ur Kernphysik, Johannes
Gutenberg-Universit\"at, D-55099 Mainz, Germany}}

\begin{abstract}
%The complex-mass scheme (CMS) provides a consistent framework of dealing with unstable states
%in quantum field theory.
%This approach has proven successful in various applications.
%As an application of the CMS in chiral effective field theory we consider the form factor of the pion in the
%time-like region and the magnetic moment of the Roper resonance.

The complex-mass scheme (CMS) provides a consistent framework for dealing
with unstable particles in quantum field theory and has been
successfully applied to various loop calculations. As applications of
the CMS in chiral effective field theory we consider the form
factor of the pion in the time-like region and the magnetic moment of
the Roper resonance.
\end{abstract}

\maketitle

%%%%%%%%%%%%%%%%%%%%%%%%%%%%%%%%%%%%%%%%%%%%
%% MAINMATTER
%%%%%%%%%%%%%%%%%%%%%%%%%%%%%%%%%%%%%%%%%%%%

\section{introduction}

	Chiral perturbation theory (ChPT) is an established low energy effective field theory of QCD
in the vacuum sector \cite{Weinberg:1978kz,Gasser:1984yg}.
	While the problem of including the nucleon and the $\Delta$ in this approach has been
resolved \cite{Scherer:2009bt}, the treatment of other
states is more complicated.
	A solution to the problem of inclusion of unstable particles in chiral EFT
is provided by the complex-mass scheme (CMS)
\cite{Stuart:1990,Denner:1999gp}.
   As an application of CMS the mass and
the width of the $\rho$ meson and the Roper resonance have been considered \cite{Djukanovic:2009zn,Djukanovic:2010zz}.
	In this contribution we present the results for the form factor of the pion in the time-like region
up to $q^2\sim 1\, {\rm GeV^2}$ and the magnetic moment of the Roper resonance up to ${\cal
O}(q^3)$.

\section{Pion form factor}

	We start with the effective Lagrangian of pions and $\rho$ mesons given as \cite{Djukanovic:2009zn}
\begin{eqnarray}
{\cal L} & = & \frac{F^2}{4}\,{\rm Tr} \left[D_\mu U \left(D^\mu
U\right)^\dagger\right]+\frac{F^2}{4}\,{\rm Tr} \left[ \chi
U^\dagger+U \chi^\dagger\right] \nonumber\\ &+& \frac{M^2 +
c_{x}{\rm Tr}\left[ \chi_+\right]/4}{g^2}\,{\rm Tr}\left[\left( g
\rho^\mu-i \Gamma^\mu\right)\left(g \rho_{\mu}-i \Gamma_\mu
\right)\right]\nonumber\\
&-&\frac{1}{2}\,{\rm
Tr}\left[\rho_{\mu\nu}\rho^{\mu\nu}\right]+i\,d_x {\rm
Tr}\left[\rho^{\mu\nu}\Gamma_{\mu\nu}\right]
- \frac{\sqrt{2}}{2}f_V{\rm
Tr}\left\{\rho_{\mu\nu}f^{\mu\nu}\right\}+\cdots ,
\label{LagrangianVCh}
\end{eqnarray}
where dots stand for terms with more derivatives/fields and
\begin{eqnarray}
U(x) & = & u^2(x) = \exp\left[\frac{i
\Phi(x)}{F}\right]\,,\
\ \
D_\mu A  =  \partial_\mu A -i v_\mu A+i A v_\mu\,,\nonumber
\\
%r_\mu & = & v_\mu+a_\mu\,,\nonumber
%\\
%l_\mu & = & v_\mu-a_\mu\,, \nonumber
%\\
\chi_+ & = & M^2(U^\dagger + U), \
\rho_{\mu}  =  \rho_{\mu}^a \,\frac{\tau^a}{2},\ \Gamma_{\mu\nu} =
\partial_\mu \Gamma_\nu-\partial_\nu \Gamma_\mu
+[\Gamma_\mu,\Gamma_\nu]
,\nonumber
\\
% V^{\mu\nu} & = & \nabla^\mu V^\nu-\nabla^\nu V^\mu\,,\nonumber\\
% \nabla_\mu V_\nu & = & \partial_\mu
% V_\nu+[\Gamma_\mu,V_\nu]\nonumber
% \\
\Gamma_\mu &= & \frac{1}{2}\,\biggl[ u^\dagger\partial_\mu u+u
\partial_\mu u^\dagger
- i\,\left( u^\dagger v_\mu u +u v_\mu u^\dagger\right)\biggr]\,,
\nonumber\\
\rho^{\mu\nu}  & = &
\partial^\mu\rho^\nu-\partial^\nu\rho^\mu - i
g\left[\rho^\mu,\rho^\nu\right]
\,,
\nonumber
\\
f_{\mu\nu} & = & u (\partial_\mu v_\nu - \partial_\nu v_\mu) u^\dagger + u^\dagger
(\partial_\mu v_\nu - \partial_\nu v_\mu) u
%\nonumber
%\\
%{\cal V}_{\mu\nu} & = & \rho_{\mu\nu}
%-\frac{i}{g_0}\,\Gamma_{\mu\nu}
\,.
\label{somedefinitions}
\end{eqnarray}
Here $F$ is the pion-decay constant in the chiral limit, $M$ is the pion mass at leading-order,
$v_\mu$ and $a_\mu$ are the external vector and axial vector
fields and $M^2=2\,g^2 F^2$ \cite{Kawarabayashi:1966kd,Riazuddin:sw,Djukanovic:2004mm}.
%For the electromagnetic interaction we have $v_\mu=-e\,\tau^3 A_\mu/2$.

 In the following we use the CMS  as renormalization scheme.
%	We choose the renormalized mass of the vector meson as the pole of the
%propagator in the chiral limit $M_R^2=(M_\rho- i \Gamma/2)^2$.
    To determine the chiral order of a diagram we consider all
possible flows of the external momenta through the
internal lines and determine the chiral order for each
flow:
	Pion propagators which do not carry large momenta are counted
as ${\cal O}(q^{-2})$, and those carrying large
momenta - as ${\cal O}(q^{0})$, vector meson propagators
not carrying large momenta - as ${\cal O}(q^{0})$,
and those carrying large momenta - as ${\cal O}(q^{-1})$. The pion mass counts as ${\cal
O}(q^{1})$, the vector meson mass - as ${\cal O}(q^{0})$
and the width - as ${\cal O}(q^{1})$. Vertices generated
by  ${\cal L}_\pi^{(n)}$ count as  ${\cal O}(q^n)$. On the other
hand derivatives acting on heavy vector mesons are counted as
${\cal O}(q^0)$. The chiral order of the diagram is
given by the smallest amongst the orders of different flows.

\begin{figure}
\includegraphics[height=.21\textheight]{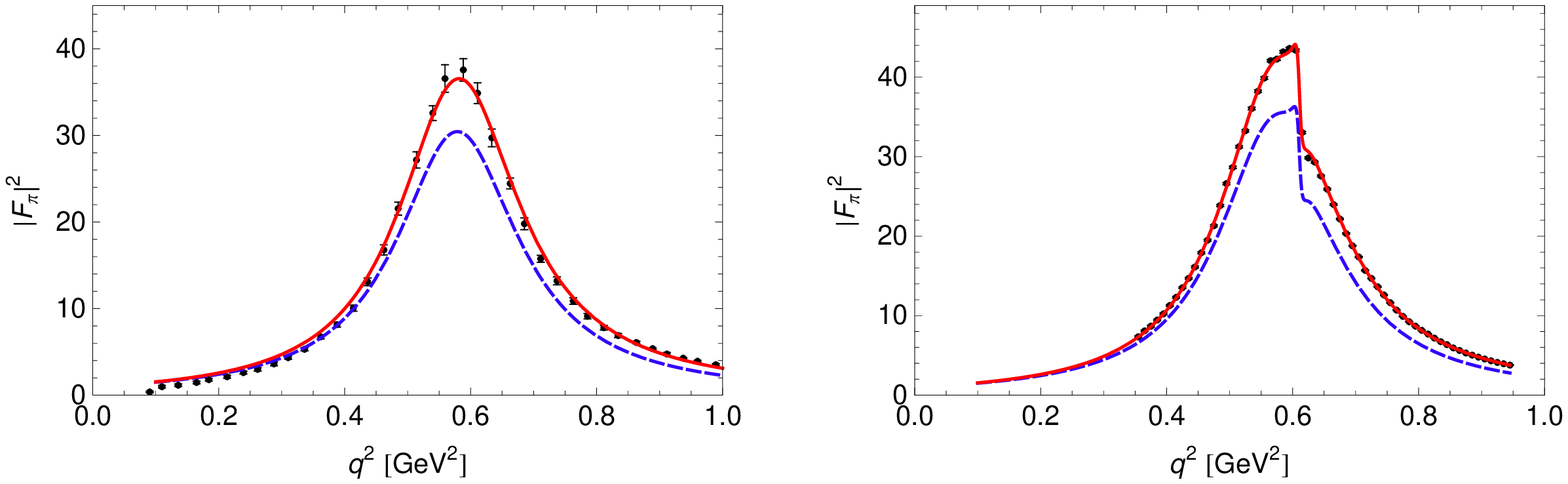}
\caption{\label{VF2:fig} Electromagnetic form factor of the pion. Left panel - extracted from
$\tau^-\rightarrow \nu_\tau \pi^- \pi^0$ decay, data - from  Ref.~\cite{Schael:2005am}.
Right panel - extracted from $e^+ e^-\rightarrow \pi^+ \pi^-$,
data - from Ref.~\cite{:2008en}.
The dashed lines show the results of the tree diagrams and the solid lines of the
tree plus one-loop diagrams respectively.}
\end{figure}

We have calculated tree and one-loop order contributions to the pion form factor $F(q^2)$.
After renormalization $F(q^2)$ is dominated by tree order diagrams with vector meson exchange.
	We fitted the available parameters of the effective Lagrangian to the
pion form factor extracted from the $\tau^-\rightarrow \nu_\tau \pi^- \pi^0$ decay. We also calculated the form factor at tree order for the same values of the parameters by switching off the loop contributions. Our results %together with the experimental data
are plotted in the left panel of Fig.~\ref{VF2:fig}.

%\newpage

	The $\rho^0-\omega-A$ mixing plays an important role in
describing the pion form factor extracted from $e^+ e^-\to \pi^+\pi^-$. Within the QFT formalism the above mixing
is taken into account by solving the system of coupled equations for dressed propagators.
	We take into account the $\rho-\omega-A$ mixing only at tree order. However we allow
the mixing parameters to become complex. %, thus incorporating the contributions of loop diagrams in the renormalization of the mixing parameters.
By fitting the mixing parameters to the data we obtain the results for the pion form factor
plotted in the right panel of Fig.~\ref{VF2:fig} together with the experimental data and the form factor at tree
order for the same values of the parameters.

\section{Magnetic moment of the Roper resonance}

The leading-order effective Lagrangian relevant for the calculation of the magnetic moment
 at ${\cal O}(q^3)$ is given by
\begin{eqnarray}
\mathcal{L}_{0} & = & \bar{N}\,(i
D\hspace{-.65em}/\hspace{.1em}-m_{N})N + \bar{R}(i
D\hspace{-.65em}/\hspace{.1em}-m_{R}) R \nonumber\\
&-&
 \bar\Psi_\mu\xi^{\frac{3}{2}}
\biggl[(i {D\hspace{-.65em}/\hspace{.1em}}-m_{\Delta })\,g^{\mu\nu}
-i\,(\gamma^{\mu}D^{\nu}+\gamma^{\nu}D^{\mu})+i\,\gamma^{\mu}
{D\hspace{-.65em}/\hspace{.3em}}\gamma^{\nu} + m_{\Delta
}\,\gamma^{\mu}\gamma^{\nu}\biggr]\xi^{\frac{3}{2}} \Psi_\nu .
\end{eqnarray}
   Here, $N$ and $R$ denote nucleon and Roper resonance fields
and $\Psi_\nu$ is the Rarita-Schwinger field of the $\Delta$ resonance.
%$\xi^{\frac{3}{2}}$ is the isospin projector.
   The covariant derivatives are defined as
\begin{eqnarray}
D_\mu H & = & \left( \partial_\mu + \Gamma_\mu-i\,v_\mu ^{(s)}\right) H\,, \nonumber\\
\left(D_\mu\Psi\right)_{\nu,i} & = &
\partial_\mu\Psi_{\nu,i}-2\,i\,\epsilon_{ijk}\Gamma_{\mu,k} \Psi_{\nu,j}+\Gamma_\mu\Psi_{\nu,i}
-i\,v_\mu^{(s)}\Psi_{\nu,i}\,,%\nonumber\\
%\Gamma_\mu & = &
%\frac{1}{2}\,\left[u^\dagger \partial_\mu u +u
%\partial_\mu u^\dagger-i\,\left( u^\dagger v_\mu u+u v_\mu u^\dagger
%\right)\right]=\tau_k\Gamma_{\mu,k},
\label{cders}
\end{eqnarray}
where $H$ stands either for the nucleon or the Roper resonance and $u=\sqrt{U}$.

%   The lowest-order Goldstone-boson Lagrangian ${\cal L}_\pi$ can be extracted from Eq.~(\ref{LagrangianVCh}) and
%the
Further interaction terms $\mathcal{L}_{R}$, $\mathcal{L}_{NR}$, and $\mathcal{L}_{\Delta R}$
are constructed in analogy to Ref.~\cite{Borasoy:2006fk}:
\begin{eqnarray}
{\cal L}_R^{(1)} & = & \frac{g_R}{2}\,\bar R \gamma^\mu\gamma_5 u_\mu
R\,, \ \
{\cal L}_R^{(2)}  =  \bar R\left[  \frac{c_{6}^*}{2}
\,f^+_{\mu\nu}+\frac{c_{7}^*}{2} \,v^{(s)}_{\mu\nu}
\right]\,\sigma^{\mu\nu}R+\cdots \,,\nonumber\\
{\cal L}_R^{(3)} & = & \frac{i}{2}\,d_{6}^* \bar R\left[
D^\mu,f^+_{\mu\nu}\right]\,D^\nu R+{\rm h.c.}+ 2\,i\,d_{7}^* \bar
R\left(
\partial^\mu v^{(s)}_{\mu\nu}\right)\,D^\nu R+{\rm h.c.}+ \cdots \,,\nonumber\\
{\cal L}_{N R}^{(1)} & = & \frac{g_{N R}}{2}\,\bar R
\gamma^\mu\gamma_5 u_\mu N+ {\rm h.c.},\ \
{\cal L}_{\Delta R}^{(1)} = - g_{\Delta R} \,\bar{\Psi}_{\mu}
\,\xi^{\frac{3}{2}} \,(g^{\mu\nu}
+\tilde{z}\,\gamma^{\mu}\gamma^{\nu})\, u_{\nu}\, R + {\rm h.c.}\,,\nonumber\\
u_\mu & = & i \left[u^\dagger \partial_\mu u -u \partial_\mu
u^\dagger-i\,\left( u^\dagger v_\mu u-u v_\mu u^\dagger
\right)\right], \nonumber\\
v_{\mu\nu}^{(s)} & = & \partial_\mu v^{(s)}_\nu - \partial_\nu v^{(s)}_\mu,
f_{\mu\nu}^{+}  =  u f_{\mu\nu} u^\dagger +u^\dagger f_{\mu\nu} u,
f_{\mu\nu}  =  \partial_\mu v_\nu - \partial_\nu v_\mu-i
[v_\mu,v_\nu] \label{bbks}
\end{eqnarray}
and $g_R$, $g_{N R}$, $g_{\Delta R}$, $c_{6}^*$, $c_{7}^*$, $d_{6}^*$, $d_{7}^*$ are unknown
coupling constants and we take $\tilde z=-1$.

\begin{figure}
\includegraphics[height=.21\textheight]{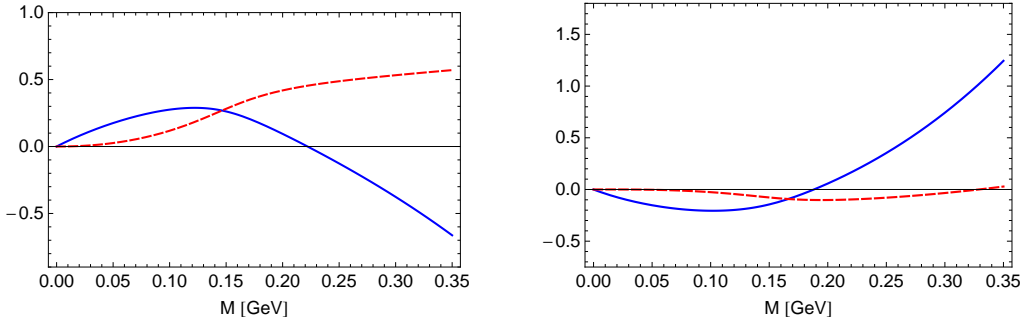}
\caption{\label{RMM:fig} One-loop contributions to the anomalous
magnetic moment of the neutral(left) and the charged(right) Roper resonances as functions of the pion mass M.
 The solid and dashed lines indicate the real and
imaginary parts, respectively.}
\end{figure}

We  renormalize loop diagrams by applying the CMS and use the standard
power counting of Ref.~\cite{Weinberg:1991um}.
    Following Ref.~\cite{Gegelia:2009py}, we parameterize the vertex function as
\begin{equation}
\sqrt{Z_R}\,\bar w^i(p_f) \Gamma^\mu(p_f,p_i) w^j(p_i)\sqrt{Z_R}=
\bar w^i(p_f)\left[\,\gamma^\mu\, F_1(q^2)  + \frac{i
\,\sigma^{\mu\nu}\, q_\nu}{2\,m_N}\,F_2(q^2)\,\right]w^j(p_i) \,,
\label{FF}
\end{equation}
where $q=p_f-p_i$, $m_N$ is the mass of the nucleon, and
$\bar w^i$ and $w^j$ denote ''Dirac spinors'' with complex masses
$z$.
    To ${\cal O}(q^3)$ the vertex function $\Gamma^\mu(p_f,p_i)$
obtains contributions from three tree diagrams
%, shown in Fig.~\ref{roperFFTree:fig},
and fourteen loop diagrams.
% shown in Fig.~\ref{MassWidth:fig}.
   We obtain $F_1(0)=(1+\tau_3)/2$ and the power-counting-violating loop contributions to
the magnetic form factor are absorbed in the couplings $c_{6}^*$ and $c_7^*$.
The tree order result for $\kappa_R= F_2(0)$ is given by
\begin{equation}
\kappa_R^{\rm tree} = 2\,m_N
\left(\frac{c_7^*}{2}+\tau_3\,c_6^*\right).
\end{equation}
   The estimated value of  the loop contributions to the anomalous magnetic moment is  %of the Roper
%we substitute $F= 0.092\,\, {\rm GeV}$, $M = 0.140\,\,
%{\rm GeV}$, $m=0.940\,\, {\rm GeV}$, $z_{\Delta \chi} = (1.210 -
%0.100\,i/2)\,\, {\rm GeV}$, $z_\chi = (1.365-0.190\,i/2)\,\, {\rm
%GeV}$, $\mu = 1\,\, {\rm GeV}$, $g_R=1$, $g_{\Delta R} = 1$,
%$g_{NR}= 0.45$ \cite{Borasoy:2006fk} and obtain
\begin{equation}
\kappa_R = (0.055+ 0.090\,i)-(0.223+ 0.156\,i)
   \,\tau_3 \label{munumeric}
\end{equation}
and Fig.~\ref{RMM:fig} shows the loop contribution as a function of the lowest-order pion mass $M$.

\section{summary}

	We have presented the results for the form factor of the pion and the magnetic moment of the Roper resonance
in the framework of chiral EFT with resonances. % using the CMS.
%Within this renormalization scheme the considered EFT possesses a consistent power counting.
Taking into account the $\rho^0-\omega-A$ mixing at tree level and by fitting the parameters of the
Lagrangian a satisfactory description has been obtained for form factors obtained both from
$\tau^-\rightarrow \nu_\tau \pi^- \pi^0$ and $e^+ e^-\to \pi^+\pi^-$. %More work
%has to be done to incorporate the isospin symmetry breaking
%effects systematically.
While no measured value of the magnetic moment of the Roper resonance is available, our expressions could be used in lattice
extrapolations.

%\

\begin{theacknowledgments}
%The authors thank S.~Leupold for helpful communication.
   This work was supported by the Deutsche
Forschungsgemeinschaft (SFB 443).
\end{theacknowledgments}

%%%%%%%%%%%%%%%%%%%%%%%%%%%%%%%%%%%%%%%%%%%%%%%%
%% BACKMATTER
%%%%%%%%%%%%%%%%%%%%%%%%%%%%%%%%%%%%%%%%%%%%%%%%



\bibliographystyle{aipproc}   % if natbib is available

%\bibliography{sample}

\IfFileExists{\jobname.bbl}{}
 {\typeout{}
  \typeout{******************************************}
  \typeout{** Please run "bibtex \jobname" to optain}
  \typeout{** the bibliography and then re-run LaTeX}
  \typeout{** twice to fix the references!}
  \typeout{******************************************}
  \typeout{}
 }

\end{document}